\documentclass[biber]{nowbook}

\usepackage[english]{babel}

\usepackage{longtable}
\usepackage{wrapfig}
\usepackage{graphicx}

\usepackage{amsmath,verbatim,xcolor}

\usepackage{amssymb}
\usepackage{enumitem}
\usepackage{listings}
\usepackage{authblk}
\usepackage{textcomp}

\usepackage[noend]{algpseudocode}
\usepackage{algorithm}
\usepackage{url}
\usepackage{bbm}

\usepackage[T1]{fontenc}

\usepackage{tikz}
\usetikzlibrary{shapes.geometric}
\usetikzlibrary{positioning}

\usepackage{caption}
\usepackage{subcaption}
\usepackage{multirow}

\newcommand{\reals}{{\mbox{\bf R}}}

\newcommand{\BEQ}{\begin{equation}}
\newcommand{\EEQ}{\end{equation}}
\newcommand{\BEAS}{\begin{eqnarray*}}
\newcommand{\EEAS}{\end{eqnarray*}}
\newcommand{\BEA}{\begin{eqnarray}}
\newcommand{\EEA}{\end{eqnarray}}    

\newcommand{\BBM}{\left[\begin{matrix}}
\newcommand{\EBM}{\end{matrix}\right]}

\newcommand{\BIT}{\begin{itemize}}
\newcommand{\EIT}{\end{itemize}}
\newcommand{\BNUM}{\begin{enumerate}}
\newcommand{\ENUM}{\end{enumerate}}

\newcommand{\argmin}{\mathop{\rm argmin}}

\newcommand{\eg}{{\it e.g.}}
\newcommand{\ie}{{\it i.e.}}

\newcommand{\PAR}[1]{\paragraph{#1.}}

\makeatletter
\let\nowfnt@unnumbered@chapterhead\nowfnt@unnumberedboxed@chapterhead
\makeatother

\graphicspath{ {./figures/} }

\begin{document}








\mainmatter

\setcounter{chapter}{1}

\chapter*{Iteratively Saturated Kalman Filtering}

\begin{center}
Alan Yang\quad Stephen Boyd
\end{center}
  
\section*{Abstract}

The Kalman filter (KF) provides optimal recursive state estimates for
linear-Gaussian systems and underpins applications in control, signal
processing, and others. However, it is vulnerable to outliers in the
measurements and process noise. We introduce the iteratively saturated Kalman
filter (ISKF), which is derived as a scaled gradient method for solving a convex
robust estimation problem. It achieves outlier robustness while preserving the
KF's low per-step cost and implementation simplicity, since in practice it
typically requires only one or two iterations to achieve good performance. The
ISKF also admits a steady-state variant that, like the standard steady-state KF,
does not require linear system solves in each time step, making it well-suited
for real-time systems.

\section{Introduction}

The Kalman filter is the prevalent tool for state estimation, prized for its
simplicity, low computational cost, and optimality for linear-Gaussian systems.
It has found extensive use in many fields, including control, signal processing,
robotics, navigation, neural interface systems, and econometrics
\cite{Kalman1960,MalikTBH2010,SmetsW2007,Huber2022}. 
Despite its popularity, the KF is notoriously vulnerable to outliers in the
measurements and process noise in the dynamics \cite{MasreliezM1977}.
Measurement outliers may arise from occasional sensor malfunctions, while
process noise outliers can result from sudden shocks to the system or unmodeled
dynamics. 

In this work, we propose the iteratively saturated Kalman filter, which is a
modification of the standard KF's update (or correction) step. It iterates a
modified KF update step, in which a saturating nonlinearity is applied to
compensate for both measurement and process noise outliers. The method is
derived as a scaled gradient method \cite{Davidon1959,FletcherR1963} for solving
a particular convex robust estimation problem involving the Huber function.
Since the ISKF typically requires only one or two iterations to achieve good
performance, it retains the standard KF's ease of implementation and per-step
cost.

A key advantage of the ISKF is its steady-state variant, which matches the
computational efficiency of the steady-state KF. Whereas the full KF must update
its covariance estimate at each step, incurring matrix-matrix multiplications
and linear solves, the steady-state KF uses precomputable gain matrices and
requires only matrix-vector multiplications and vector additions.  Our
steady-state ISKF inherits this low per-step cost while compensating for both
measurement and process-noise outliers. In contrast, existing robust KF
extensions either lack robustness to process-noise outliers or rely on full
covariance estimates in each step.

The rest of the paper is organized as follows. We review prior work in
\S\ref{s-prior-work}. The system model is given in \S\ref{s-system-model}, and
the iteratively saturated Kalman filter is introduced in \S\ref{s-hkf}. We
derive the filter as a scaled gradient method in \S\ref{s-hkf-scaled-gradient}.
Finally, we present numerical experiments in \S\ref{s-experiments} and conclude
in \S\ref{s-conclusion}.

\section{Prior work}\label{s-prior-work}

There is a large body of work on modifying the KF to be robust to outliers
without sacrificing its computational efficiency. Many of these heuristics
involve modifying the covariance estimate in each KF step, by scaling the
measurement noise covariance matrix or the process and prior covariance matrices
when outliers are detected \cite{DurovicK1999,Chang2014,DuranMartinAS2024}. The
idea is that if an outlier is detected, the corresponding covariance should be
scaled up to account for the increased uncertainty. Some variations of this idea
were derived by replacing the Gaussian distribution used by the KF with
heavy-tailed distributions \cite{TingTS2007,AgamennoniNN2011}. Others are
derived by Huberizing the quadratic costs used by the KF
\cite{Huber1964,MasreliezM1977,CipraR1991,KovacevicD1992,DurovicK1999,MattingleyB2012}.

Yet others have proposed combining the KF with inlier detection methods such as
RANSAC \cite{Cantzler1981,VedaldiJFS2005}. Several of the robust KF methods have
also been extended to nonlinear systems \cite{PicheSH2012,Karlgaard2015}.

Our method generalizes the saturated KF \cite{FangHHW2018} by compensating for
process noise outliers in addition to measurement outliers. In the single-step
case, our method is almost equivalent to the saturated KF, but uses a different
saturation function.

Besides modified KF methods, we also mention particle filtering methods, which
have been applied to handle outliers \cite{BoustatiADJ2020}, although they tend
to be computationally expensive.

\section{System model}\label{s-system-model}
We consider a linear time-invariant dynamical system that evolves according to
\BEQ\label{e-system-model}
x_{t+1} = A x_t + w_t, \quad y_t = C x_t + v_t,
\quad t = 0,1,\ldots,
\EEQ
where $t$ denotes time or epoch, $x_t\in\reals^n$ is the state, and
$y_t\in\reals^p$ is the output measurement. The matrix $A\in\reals^{n\times n}$
is the state dynamics matrix, and $C\in\reals^{p\times n}$ is the output matrix.
We assume that the dynamics matrix $A$ and the output matrix $C$ are known. The
dynamics are driven by the process noise $w_t\in\reals^n$, and the outputs are
influenced by the measurement noise $v_t\in\reals^p$. We assume that the initial
state $x_0$ is Gaussian, with $x_0\sim\mathcal N(0, X_0)$.

\PAR{Linear Gaussian model}

In the classical model, the process noise $w_t\in\reals^n$ and the measurement
noise $v_t\in\reals^p$ are 
Gaussian with
\[
w_t \sim\mathcal N(0, W),\quad v_t \sim \mathcal N(0, V),
\]
where $W$ is the known positive semidefinite (PSD) process noise covariance
(which can be degenerate, \ie, singular) and $V$ is the known positive definite
(PD) measurement noise covariance. We assume that the initial state $x_0$, the
process noise $w_t$, and the measurement noise $v_t$ are independent and
identically distributed (IID).
In this case, the state and measurements are jointly Gaussian, and the Kalman
filter \cite{Kalman1960} gives the optimal estimate of the state, both in the
minimum mean squared error (MMSE) and maximum a posteriori (MAP) sense.

\PAR{Outlier model}

In this work, we consider a model where the process and measurement noises are
typically Gaussian, but may occasionally be corrupted by outliers. We consider
the model
\BEA\label{e-outlier-model}
w_t = F(\tilde w_t + s_t), \quad v_t = G(\tilde v_t + o_t),
\EEA
where $F\in\reals^{n\times m}$ and $G\in\reals^{p\times p}$ are known matrices,
and $\tilde w_t\in\reals^m$ and $\tilde v_t\in\reals^p$ are zero-mean whitened
Gaussian noises with $\tilde w_t\sim\mathcal N(0,I)$ and $\tilde v_t\sim\mathcal
N(0,I)$. The additional terms $s_t\in\reals^n$ and $o_t\in\reals^p$ are sparse
outlier terms which are zero for most times $t$.

The process noise outliers $s_t$ can result from system shocks or unmodeled
dynamics, while measurement outliers $o_t$ can arise from sensor malfunctions or
environmental disturbances. In the absence of outliers, \ie, when $s_t = 0$ and
$o_t = 0$ for all $t$, the system reduces to the linear Gaussian model, with
$W=FF^T$ and $V=GG^T$.

\PAR{Extensions}

Several extensions of the system model are readily handled. The model can be
modified to handle known control inputs, process and measurement noises $w_t$
and $v_t$ with nonzero mean, and correlation between $w_t$ and $v_t$. We may
also consider the time-varying case, where $A$, $C$, $W$, and $V$ are allowed to
vary with $t$.

\section{Iteratively saturated Kalman filtering}\label{s-hkf}

Given a sequence of measurements $y_1,\ldots,y_t$, our goal is to recursively
estimate both the state $x_t$ and its covariance $P_t$. At each time step $t$,
we update our previous estimates $\hat x_{t-1\mid t-1}$ and $P_{t-1\mid t-1}$ to
obtain new estimates $\hat x_{t\mid t}$ and $P_{t\mid t}$. We assume in the
following that $(A,C)$ is detectable and $(A,W^{1/2})$ is stabilizable
\cite{Simon2006}. Here, $W^{1/2}$ denotes any matrix such that $(W^{1/2})^T
W^{1/2} = W$. Such a matrix can be found from the eigenvalue decomposition of
$W$, or when $W$ is PD, via Cholesky factorization.

Our estimator starts with the standard Kalman filter prediction step
\BEA
\hat x_{t\mid t-1} &=& A \hat x_{t-1\mid t-1} \label{e-hkf-pred-mean} \\
P_{t\mid t-1} &=& A P_{t-1\mid t-1} A^T + W. \label{e-hkf-pred-cov}
\EEA
The update step is then given by the iteration
\BEA
\hat x^0 &=& \hat x_{t\mid t-1} \\
\hat x^k &=& \hat x^{k-1}
+ K_t \sigma(y_t - C\hat x^{k-1}) 
+ (I - K_t C)\rho_t(\hat x^0 - \hat x^{k-1}), \label{e-hkf-update-mean}
\EEA
for $k = 1,\ldots, \tilde k$, where $\tilde k$ denotes the number of iterations.
We then take $\hat x_{t\mid t} = \hat x^{\tilde k}$.
The nonlinear functions
\[
\rho_t(z) = \min\left(1, \frac{\lambda_x}{\|P_{t\mid t-1}^{-1/2}z\|_2}\right) z, \quad
\sigma(z) = \min\left(1, \frac{\lambda_y}{\|V^{-1/2}z\|_2}\right) z,
\]
saturate their inputs at threshold values $\lambda_x$ and $\lambda_y$,
respectively, and $K_t$ is the Kalman gain matrix satisfying the standard Kalman
filter covariance update equations
\BEA
K_t &=& P_{t\mid t-1} C^T (C P_{t\mid t-1} C^T + V)^{-1}
\label{e-hkf-gain} \\
P_{t\mid t} &=& (I - K_t C) P_{t\mid t-1}.
\label{e-hkf-update-cov}
\EEA

\PAR{Comments}

Our filter compensates for outliers in the measurements by reducing the effect
of outliers in the measurement noise on the state estimate by attenuating the
magnitude of the \emph{innovation} $y_t - C \hat x_{t\mid t-1}$ when
\[
\|V^{-1/2}(y_t - C \hat x_{t\mid t-1})\|_2
\]
is large, \ie, unlikely under the Gaussian noise model. Similarly, our filter
compensates for outliers in the process noise by attenuating the magnitude of
the proximity to the predicted state $\hat x_{t\mid t-1}$.

\PAR{Single-step case}

When $\tilde k=1$, the update step \eqref{e-hkf-update-mean} is simply
\[
\hat x_{t\mid t} = \hat x_{t\mid t-1} + K_t\sigma(y_t - C \hat x_{t\mid t-1}).
\]
In this case, the ISKF does not compensate for outliers in the process noise,
only in the measurements. It closely resembles the standard KF, except that the
innovation $y_t - C \hat x_{t\mid t-1}$ is attenuated by the function $\sigma$.
The single-step ISKF is very similar to the saturated KF \cite{FangHHW2018}, but
uses a different saturation function.

\PAR{Outlier-free case}

In the absence of detected outliers, our state estimate $\hat x_{t\mid t}$
satisfies
\[
\|V^{-1/2}(y_t - C \hat x_{t\mid t})\|_2 \le \lambda_y, \quad
\|P_{t\mid t-1}^{-1/2}(\hat x_{t\mid t} - \hat x_{t\mid t-1})\|_2 \le \lambda_x,
\]
and the saturation functions $\rho_t$ and $\sigma$ reduce to the identity
function. In this case, the ISKF is equivalent to the standard KF
\[
\hat x^{\textrm{kf}}_{t\mid t} = \hat x_{t\mid t-1} + K_t(y_t - C \hat x_{t\mid t-1}),
\]
with $\hat x^k = \hat x^{\textrm{kf}}_{t\mid t}$ for all $k$.

\PAR{Computational cost}

At each time step, the ISKF costs $O(\tilde k(n^3 + p^3 + np))$ floating-point
operations (FLOPS) online, dominated by matrix-matrix products and solving a
linear system. For $\tilde k$ fixed and small, this is comparable to the cost of
the KF. In our experiments, we found that values between 1 and 5 were effective.

\PAR{Steady-state case}
A key property of the Kalman filter recursion is that the covariance update steps
\eqref{e-hkf-pred-cov} and \eqref{e-hkf-update-cov} are independent of the
measurements. This means we can compute $P_{t\mid t}$ offline, before processing
any measurements. Furthermore, as $t\to\infty$, the covariance matrices
$P_{t\mid t}$, $P_{t\mid t-1}$, and gain matrix $K_t$ converge to steady-state
values given by:
\BEAS
P &=& A P A^T + W - A P C^T (C P C^T + V)^{-1} C P A^T \\
\Sigma &=& APA^T + W \\
K &=& \Sigma C^T (C \Sigma C^T + V)^{-1}
\EEAS
respectively. In steady-state, we may dispense with the covariance and gain
update steps \eqref{e-hkf-pred-cov}, \eqref{e-hkf-gain}, and
\eqref{e-hkf-update-cov}.

This leads to the \emph{steady-state ISKF}
\BEA
\hat x^0 &=& \hat x_{t\mid t-1} \\
\hat x^k &=& \hat x^{k-1}
+ K \sigma(y_t - C\hat x^{k-1}) 
+ (I - K C)\rho(\hat x^0 - \hat x^{k-1}), \label{e-ss-hkf-update}
\EEA
for $k = 1,\ldots, \tilde k$. Like before, we take $\hat x_{t\mid t} = \hat
x^{\tilde k}$. Here, we drop the subscript $t$ on the function $\rho$, since
it is now time-invariant.

Since $K$ can be precomputed offline, the steady-state ISKF only requires
matrix-vector multiplications and vector additions online, with has cost
$O(\tilde k(n^2 + np))$ FLOPS online. Since the convergence of $P_t$ to $P$ is
in practice often quick, the ISKF estimate \eqref{e-ss-hkf-update} gives an
excellent approximation to the ISKF \eqref{e-hkf-update-mean}.

\section{Derivation as a scaled gradient method}\label{s-hkf-scaled-gradient}

In this section, we show that the ISKF can be interpreted as a scaled gradient
method for solving a convex regularized maximum a posteriori (MAP) estimation
problem in which we estimate $x_t$, $s_t$, and $o_t$ jointly. We focus on the
steady-state case for simplicity, but the discussion applies to the general case
by replacing $P$ with $P_{t\mid t}$ and $\Sigma$ with $P_{t\mid t-1}$, and
adding the subscript $t$ on $K$ and $\rho$.

\subsection{Model}

At each time step $t$, we consider the optimization problem
\[
\textrm{minimize} \; \frac 1 2
\left\|
\left[\begin{array}{c}
\Sigma^{-1/2}(x - \hat x_{t\mid t-1} - s) \\
V^{-1/2}(y_t - Cx - o)
\end{array}\right]
\right\|_2^2 + \lambda_x \|\Sigma^{-1/2} s\|_2 + \lambda_y \|V^{-1/2} o\|_2
\]
with variables $x$, $s$, and $o$. This is a convex optimization problem, and it
has the following interpretation. When the outlier terms $s$ and $o$ are known
and fixed, the first term corresponds to the negative log likelihood of the
prior distribution over $x_t$ and the measurement noise. The second and third
terms are sparse regularization terms on $s$ and $o$. When the thresholds
$\lambda_x$ and $\lambda_y$ are infinite, the optimal values of $s$ and $o$ are
zero, and the optimal value of $x$ is simply given by the (steady-state) KF
estimate $\hat x^{\textrm{kf}}_{t\mid t}$.

\PAR{Comments}

Our optimization problem is only an approximate MAP estimate, since in general
we do not have a statistical model of the outliers $s_t$ and $o_t$. Moreover, in
the presence of process noise outliers, the prior distribution over $x_t$ is no
longer necessarily Gaussian with mean $\hat x_{t\mid t-1}$ and covariance $P$.
The goal is to reduce the influence of outliers on the state estimate by
regularization, where the regularization terms are chosen such that the partial
minimizations over $s$ and $o$ have closed-form solutions which can be written
in terms of the Huber function.

\PAR{Circular Huber function}
Let the function $\varphi$ denote the partial
minimization 
\[
\varphi(a; \lambda) = 
\min_{b} \left(\frac 1 2 \|a-b\|_2^2 + \lambda \|b\|_2\right).
\]
The function $\varphi$ has a closed-form solution given by
\[
\varphi(a; \lambda) =
\begin{cases}
\frac 1 2 \|a\|_2^2 & \|a\|_2 \le \lambda, \\
\lambda(\|a\|_2 - \lambda/2) & \|a\|_2 > \lambda.
\end{cases}
\]
We refer to this function as the \emph{circular Huber function} with threshold
$\lambda$. It is a smooth convex function that is quadratic for $\|a\|_2 \le
\lambda$, and grows only linearly with the norm of $a$ for $\|a\|_2 > \lambda$.
It is equivalent to the standard Huber function, composed with the Euclidean
norm.

The estimation problem may then be written as
\[
\hat x_{t\mid t} = \argmin_x f(x),
\]
where
\BEQ\label{e-hkf-objective}
f(x) = \varphi(\Sigma^{-1/2}(x - \hat x_{t\mid t-1}); \lambda_x) +
\varphi(V^{-1/2}(y_t - Cx); \lambda_y).
\EEQ
For future reference, we observe that the Hessian satisfies $\nabla^2
\varphi(a;\delta) \leq I$, which implies that its gradient $\nabla
\varphi(a;\delta)$ has Lipschitz constant one.

\subsection{Scaled gradient method}\label{s-scaled-gradient}

We propose a scaled gradient method for minimizing \eqref{e-hkf-objective} over
$x\in\reals^n$. The method has iterates
\[
x^{k+1} = x^k - \eta M^{-1}\nabla f(x^k), \quad k=0,1,\ldots, \tilde k-1,
\]
where $\eta\in(0,2)$ is a constant step size,
\[
M = \Sigma^{-1} + C^T V^{-1} C
\]
is the scaling matrix, and the initial iterate $x^0 = \hat{x}_{t|t-1}$ is the
predict step given by \eqref{e-hkf-pred-mean}.

The gradient of $f$ is given by
\BEA\label{e-hkf-gradient}
\nabla f(x) &=& 
(\Sigma^{-1/2})^T 
\nabla \varphi\left(\Sigma^{-1/2}(x - \hat x_{t\mid t-1});
\lambda_x\right)
\nonumber \\
& & - C^T (V^{-1/2})^T \nabla \varphi\left(V^{-1/2}(y_t - C x); \lambda_y\right),
\EEA
where
\BEQ\label{e-huber-gradient}
\nabla \varphi(a; \lambda) = \begin{cases}
a & \|a\|_2 \le \lambda, \\
\lambda a/\|a\|_2 & \|a\|_2 > \lambda.
\end{cases}
\EEQ
To simplify the scaled gradient $M^{-1}\nabla f(x)$, we can use the fact that
the Kalman gain can be written as
\[
K = \Sigma C^T(C \Sigma C^T + V)^{-1} = M^{-1}C^T V^{-1},
\]
by applying the Woodbury matrix identity. Then, since $\nabla \varphi(a;
\lambda)$ is a scalar multiple of $a$, the scaled gradient becomes
\[
M^{-1}\nabla f(x) = 
-K \sigma(y_t - Cx) - (I - K C)\rho(\hat x_{t\mid t-1} - x).
\]
By setting the step size $\eta = 1$, we arrive at the ISKF update
\eqref{e-hkf-update-mean}. Note that in practice, it may be numerically
advantageous to implement the saturation functions as
\[
\rho(z) = \nabla \varphi(\Sigma^{-1/2}z; \lambda_x),\quad
\sigma(z) = \nabla \varphi(V^{-1/2}z; \lambda_y)
\]
where $\nabla \varphi$ has the form in \eqref{e-huber-gradient}, since the
division by $\|a\|_2$ only occurs when $\|a\|_2 > \lambda$.

\PAR{Choice of gradient method}
In general a gradient method would be a very poor choice for a problem that 
must be solved in real-time, since its practical convergence can vary considerably 
depending on the input data. An interior-point method, which typically takes
around 20 or so steps independent of the data, would seem like a better choice.
We propose the use of a gradient method in this specific case only because an
excellent estimate of the curvature of the function $f$ is available, which 
eliminates the need for a line search, and gives very good estimates in just a few
iterations, independent of the data.

\subsection{Convergence}
We now show that the scaled gradient method converges to a solution,
and is a strict descent method, when the iterations are not terminated.
Let $L$ be a matrix that
satisfies $LL^T = M^{-1}$, \eg, the Cholesky
factorization of $M^{-1}$. 
Let $g(z) = f(Lz)$. The scaled gradient method is then equivalent to the
iteration
\[
z^{k+1} = z^k - \eta \nabla g(z^k), \quad k=0,1,\ldots, T-1,
\]
where $x^k = Lz^k$ for all $k$. To establish convergence and the descent
property, it is sufficient to show that the gradient $\nabla g$ is Lipschitz
continuous with constant one \cite{Polyak1987}. Note that an upper bound on the
Lipschitz constant of $\nabla g$ can be found by taking $\lambda_x$ and
$\lambda_y$ to infinity, in which case $g$ is a quadratic function. In that case,
$g(z)$ is given by
\BEAS
g(z) &=& \|\Sigma^{-1/2}(Lz - \hat x_{t\mid t-1})\|_2^2 +
\|V^{-1/2}(y_t - C Lz)\|_2^2 \\
&=& \left\|\BBM \Sigma^{-1/2} \\ -V^{-1/2}C \EBM Lz - 
\BBM \Sigma^{-1/2}\hat x_{t\mid t-1} \\ -V^{-1/2}y_t\EBM\right\|_2^2.
\EEAS
Since
\[
L^T \BBM \Sigma^{-1/2} \\ -V^{-1/2}C \EBM^T \BBM \Sigma^{-1/2} \\ -V^{-1/2}C \EBM L 
= L^T M L = I,
\]
it follows that $\nabla g$ has Lipschitz constant one.

\section{Parameter selection}\label{s-parameter-selection}

The performance of the ISKF depends on the number of iterations $\tilde k$ and
the choice of parameters $\lambda_x$ and $\lambda_y$.

\PAR{Number of iterations}

In our experiments, we found that the ISKF can compensate for outliers in the
measurement noise even with $\tilde k=1$, and outliers in both the measurement
noise and process noise even with $\tilde k=2$. While small improvements can be
achieved for larger $\tilde k$, we found $\tilde k=2$ to be a good default
choice, with more iterations giving diminishing returns.

\PAR{Threshold parameters}

The parameters $\lambda_x$ and $\lambda_y$ balance robustness against outliers
and estimate bias. Larger values of $\lambda_x$ and $\lambda_y$ are ideal when
there are no outliers (with $\lambda_x = \lambda_y = \infty$ reducing to the
standard KF), while tuned values improve performance when outliers are present. 

\PAR{Step size}

The discussion in \S\ref{s-scaled-gradient} suggests that the step size $\eta$
in the scaled gradient method could be made a tunable parameter. The
(steady-state) ISKF \eqref{e-ss-hkf-update} can be modified as
\BEA
\hat x^0 &=& \hat x_{t\mid t-1} \\
\hat x^k &=& \hat x^{k-1}
+ \eta K \sigma(y_t - C\hat x^{k-1}) 
+ \eta (I - K C)\rho(\hat x^0 - \hat x^{k-1}). \label{e-ss-hkf-update-stepsize}
\EEA
However, we suggest fixing $\eta=1$ in practice. In our experiments, we found
that increasing $\eta$ to be greater than one could give marginal improvements,
but at the cost of reducing the performance of the filter when there are no
outliers present. This is because $\eta$ effectively acts as a type of gain
parameter for the filter. This is most clearly seen in the single-step case.
When $\tilde k=1$ and there are no detected outliers (the saturation function
$\sigma$ is identity), the ISKF reduces to the standard KF with gain matrix
$\eta K$. Choosing $\eta=1$ leads to a natural interpretation, since the filter
is then equivalent to the standard KF when there are no detected outliers.

\PAR{Grid search}

The parameters may be chosen via a simple grid search, given a sequence of
measurements $y_1,\ldots,y_N$ collected in the past. For each combination of
parameters, the ISKF can be run on the past data containing outliers, and the
parameter combination that best predicts the observed measurements is chosen.
For each parameter combination, we compute the RMSE of the predicted
measurements
\BEQ\label{e-meas-rmse}
\textrm{RMSE} = 
\left(\frac{1}{N}\sum_{t=1}^{N} \|y_{t} - CA\hat x_{t-1\mid t-1}\|_2^2\right)^{1/2},
\EEQ
although other metrics could be used. Note that we consider the RMSE of the
residuals $y_{t} - CA\hat x_{t-1\mid t-1}$, rather than the innovations $y_{t} -
C\hat x_{t\mid t}$, since the estimate $\hat x_{t\mid t}$ is computed as a
function of $y_{t}$.

In practice, a strategy for choosing the parameters is to first fix the number
of iterations $\tilde k$ based on the computational budget, and then search over
$\lambda_x$ and $\lambda_y$ to minimize the RMSE.

\section{Extensions and variations}

\subsection{Time-varying and non-linear systems}

When the system model \eqref{e-system-model} is time-varying, we introduce
subscripts $t$ to the matrices $A$, $C$, $W$, and $V$. The ISKF naturally extends
to this scenario, though a steady-state version typically does not exist. For
nonlinear systems, we approximate the model by linearizing around the current
state estimate, resulting in time-varying matrices $A_t$, $C_t$, $W_t$, and
$V_t$. Similar to how the EKF and UKF generalize the KF to nonlinear cases,
analogous modifications enable the ISKF to handle nonlinear systems by
iteratively updating the linearization.

\subsection{Missing measurements}

We have assumed thus far that the measurements $y_t$ are fully available at each
time $t$.  When this is not the case, the update step is replaced with
conditioning on only the \emph{known entries} of $y_t$. The most general way to
handle this is to allow for any subset of the entries of $y_t$ to be known or
unknown. This is equivalent to the case where the measurement matrix $C$ and
measurement covariance $V$ are time-varying, where only the rows of $C$ and $V$
corresponding to the available measurements are included. In the case where no
measurements are available, the update step is skipped. 

However, this approach leads to a time-varying system, so cannot be applied to
the steady-state case without increasing the online computational cost from
$O(n^2 + np)$ to $O(n^3 + p^3 + np)$. Alternatively, there are only $2^p$
possible patterns of missing measurements. If $2^p$ isn't too large, we could
precompute a different steady-state Kalman gain matrix $K$ for each pattern of
known measurements.




\section{Numerical experiments}\label{s-experiments}

In this section, we present numerical experiments evaluating the performance of
the ISKF, in comparison with the KF and other outlier-robust filters. In our
experiments, we use the steady-state form of the ISKF and the KF.

\PAR{Competing methods}

We compare the steady-state ISKF against the steady-state KF, and two
outlier-robust Kalman filter variants: the weighted observation likelihood
filter (WoLF) \cite{DuranMartinAS2024} and the Huberized KF
\cite{Huber1964,MasreliezM1977,CipraR1991,KovacevicD1992,DurovicK1999}. WoLF is
a covariance scaling method, and is only designed to reject measurement
outliers. Our implementation of the Huberized KF solves the Huber regression
problem \eqref{e-hkf-objective} using the interior-point method Clarabel
\cite{GoulartC2024}. Both WoLF and the Huberized KF have online computational
cost approximately $O(n^3 + p^3 + np)$, comparable to the full KF. In contrast,
the steady-state ISKF and steady-state KF have online cost $O(n^2 + np)$.

\PAR{Evaluation}

We evaluate the performance of each filter on a simulated test trajectory of,
using the state estimate RMSE
\BEQ\label{e-state-rmse}
\textrm{RMSE} = \left(\frac{1}{T} \sum_{t=1}^T \|x_t - \hat x_{t\mid t}\|_2^2\right)^{1/2},
\EEQ
where $x_t$ is the true state and $\hat x_{t\mid t}$ is the state estimate
produced by a filter. For the purposes of evaluation, we assume that the true
state trajectory is available. Test trajectories had lengths of $1000$ time
steps.

\PAR{Parameter selection}

We tuned the parameters of each filter using a separate simulated trajectory
than the test trajectory used for evaluation. Unlike the test trajectory, we
assumed that the true state trajectory was not available for the tuning
trajectory data. Instead, we minimized the predicted measurement RMSE
\eqref{e-meas-rmse} in our grid search. For all parameters, we considered 20
values between $0.1$ and $10$, logarithmically spaced. Like the test trajectory,
the tuning trajectory had length $1000$ time steps. The ISKF (for $\tilde k >
1$) and the Huberized KF have two tunable parameters, and WoLF has one.

\subsection{Vehicle tracking}\label{s-vehicle-example}

\PAR{System model}

The position and velocity of a vehicle in two dimensions are denoted $\xi_t \in
\reals^2$ and $\nu_t \in \reals^2$. At time $t$, we observe a noisy measurement
of the position $\xi_t$, and aim to estimate the state $x_t = (\xi_t, \nu_t)$.
The vehicle has unit mass, and is subject to a drag force $-\gamma \nu_t$ with
coefficient of friction $\gamma=0.05$. The discrete-time system with time step
$h=0.05$ is
\[
x_{t+1} = Ax_t + Bu_t, \quad y_t = Cx_t + v_t,
\]
where
\[
A = \begin{bmatrix}
1 & 0 & \left(1 - \frac{\gamma h}{2}\right) h & 0 \\
0 & 1 & 0 & \left(1 - \frac{\gamma h}{2}\right) h \\
0 & 0 & 1 - \gamma h & 0 \\
0 & 0 & 0 & 1 - \gamma h
\end{bmatrix},
\]
and
\[
B = \begin{bmatrix}
\frac{h^2}{2} & 0 \\
0 & \frac{h^2}{2} \\
h & 0 \\
0 & h
\end{bmatrix},
\quad
C = \begin{bmatrix}
1 & 0 & 0 & 0 \\
0 & 1 & 0 & 0
\end{bmatrix}.
\]
The vehicle is driven by a random applied force $u_t \in \reals^2$, which is
distributed according to
\[
\begin{cases}
\mathcal{N}(0, 10 I) & \text{with probability } 0.9, \\
\mathcal{N}(0, 100 I) & \text{with probability } 0.1,
\end{cases}
\]
The measurement noise $v_t\in\reals^2$ is distributed according to
\[
\begin{cases}
\mathcal{N}(0, 5 I) & \text{with probability } 0.9, \\
\mathcal{N}(0, 500 I) & \text{with probability } 0.1.
\end{cases}
\]
such that $10\%$ of the samples are large outliers. This fits the system model
\eqref{e-outlier-model} with $F = \sqrt{10} B$ and $G=\sqrt{5}I$.

\begin{figure}
\centering
\includegraphics[width=0.8\textwidth]{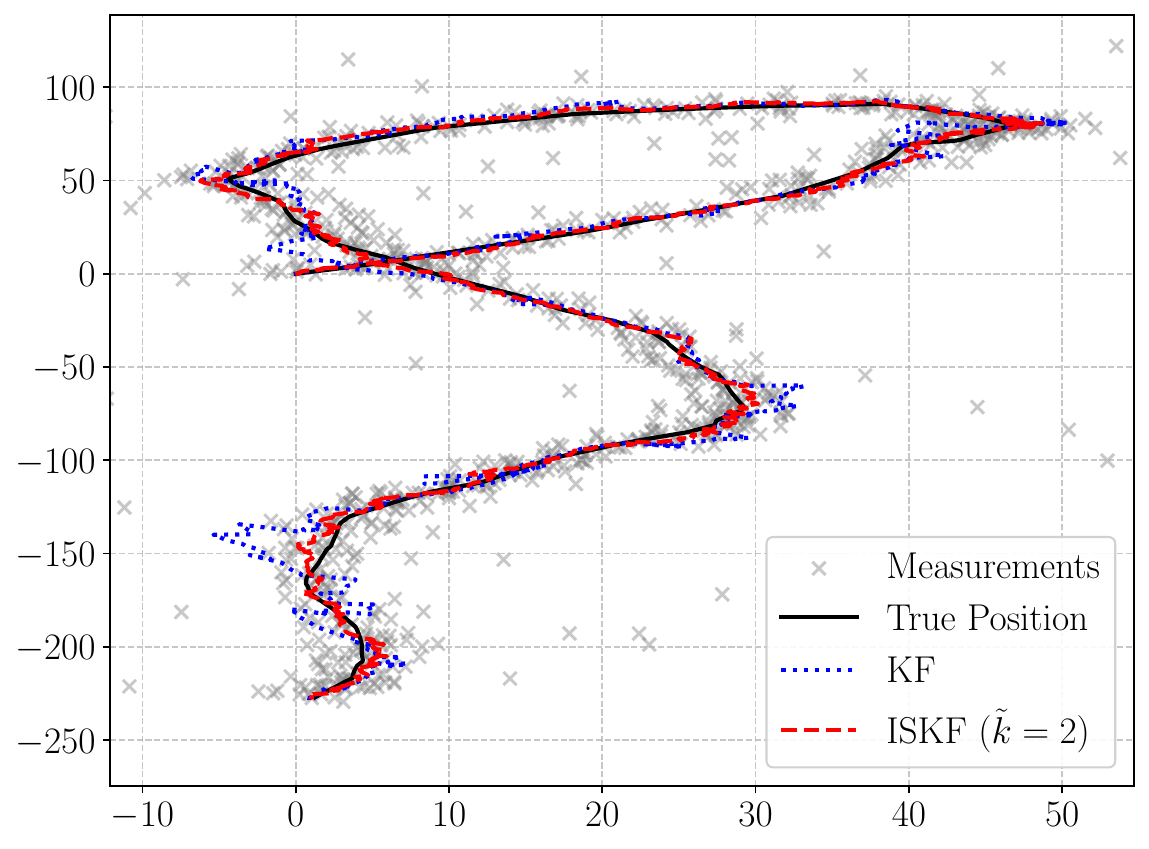}
\caption{True vehicle position over time, along with measurements and position estimates produced by the (steady-state) ISKF and KF.}
\label{f-vehicle-trajectory}
\end{figure}

\begin{figure}
\centering
\includegraphics[width=0.95\textwidth]{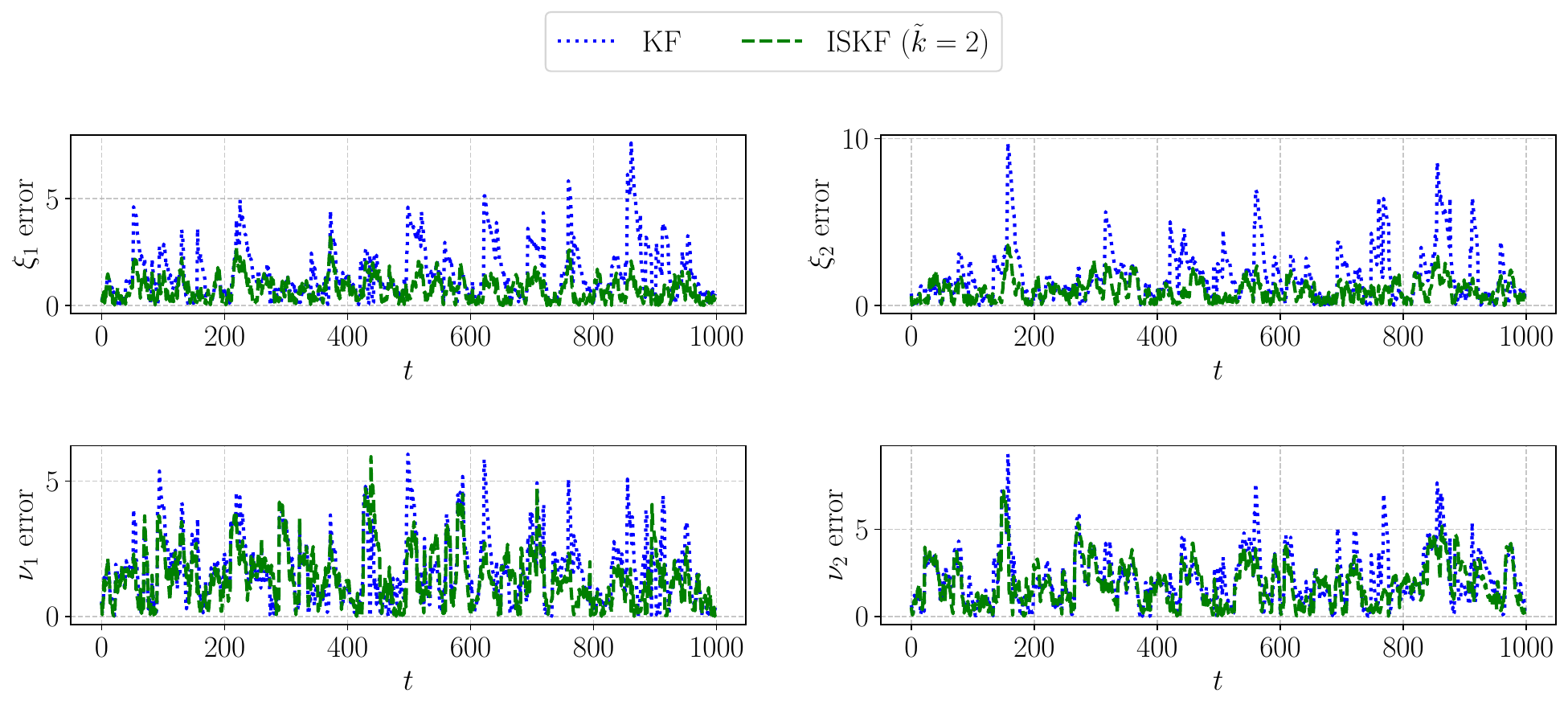}
\caption{Vehicle tracking state estimate errors (absolute values) for the KF and the two-iteration ISKF.}
\label{f-vehicle-errors}
\end{figure}

\PAR{Performance comparison}

With the number of iterations fixed at $\tilde k=2$, we selected the parameters
to be
\[
\lambda_x = 0.10, \quad \lambda_y = 1.8
\]
using the grid search procedure described in \S\ref{s-parameter-selection}. The
grid search was carried out over the predicted measurement RMSE
\eqref{e-meas-rmse} on a separate simulated trajectory of measurements. Figure
\ref{f-vehicle-trajectory} shows the true vehicle position over time, along with
measurements and position estimates produced by the (steady-state) ISKF and KF.
Figure \ref{f-vehicle-errors} shows the state estimate errors (absolute values)
for the KF and the two-iteration ISKF.

Table \ref{t-vehicle-performance} shows the state estimate RMSE
\eqref{e-state-rmse} evaluated for several filters on the same test trajectory.
The ISKF with $\tilde k=1$ achieves comparable performance to WoLF, as both are
only designed to reject measurement outliers. The (steady-state) ISKF with
$\tilde k=2$ and $\tilde k=3$ achieves better performance than WoLF, and
performs comparably to the Huberized KF, at lower computational cost.

\begin{table}[b]
\centering
\begin{tabular}{|c||c|c|}
\hline
Method & RMSE & Improvement over KF \\
\hline
\hline
KF & 4.55 & - \\
ISKF ($\tilde k = 1$) & 3.43 & 25\% \\
ISKF ($\tilde k = 2$) & 3.19 & 30\% \\
ISKF ($\tilde k = 3$) & 3.15 & 31\% \\
WoLF & 3.50 & 23\% \\
Huber KF & 3.15 & 31\% \\
\hline
\end{tabular}
\caption{State estimate RMSE comparison of several filtering methods for the
vehicle tracking example. All filters were tuned using the same data, and
evaluated on the same test data.}
\label{t-vehicle-performance}
\end{table}

\begin{figure}
\centering
\begin{subfigure}{0.45\textwidth}
\centering
\includegraphics[width=\textwidth]{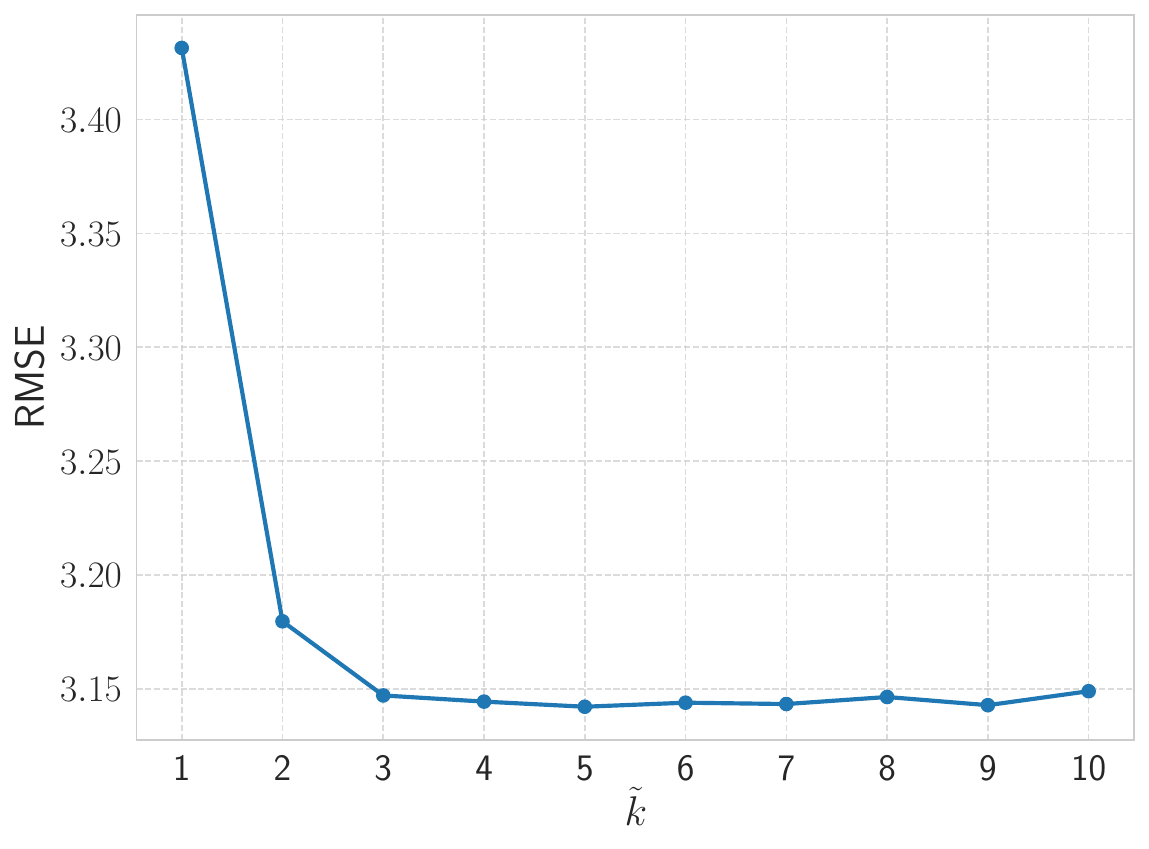}
\caption{\label{f-vehicle-iter-sweep}}
\end{subfigure}
\begin{subfigure}{0.45\textwidth}
\centering
\includegraphics[width=\textwidth]{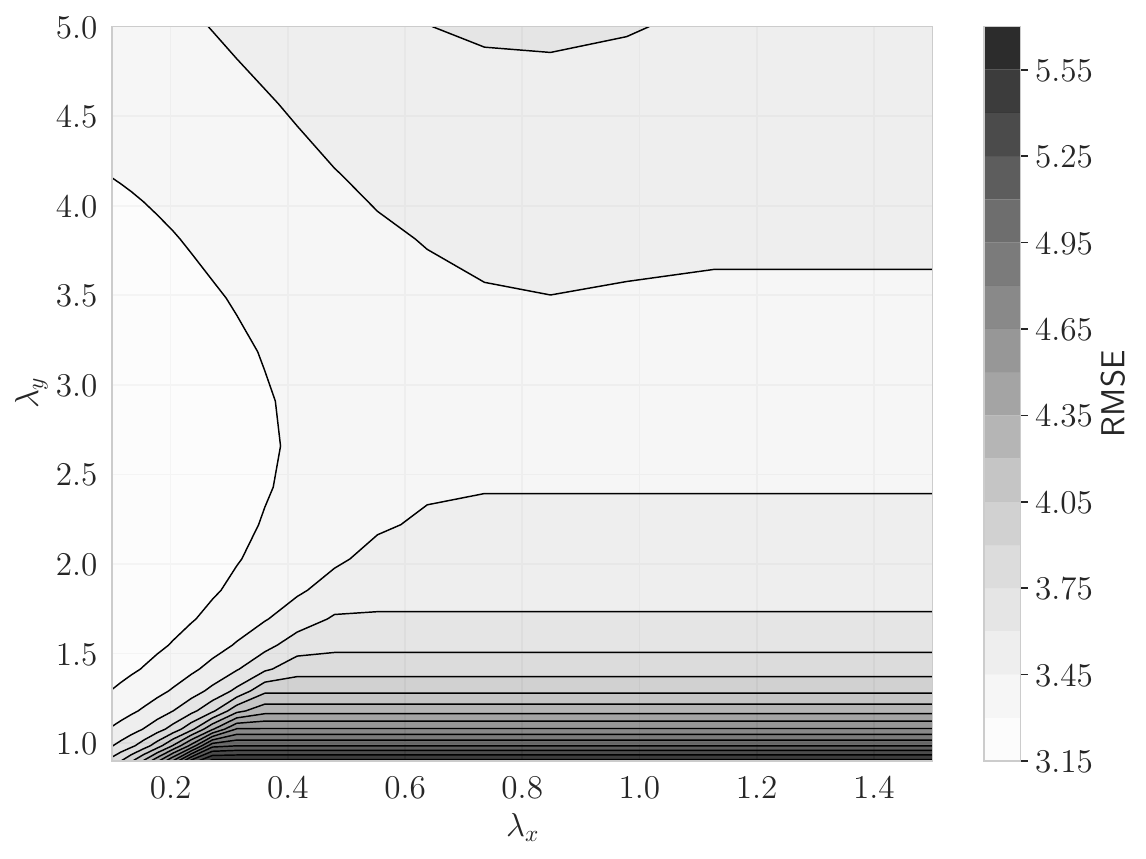}
\caption{\label{f-vehicle-contour}}
\end{subfigure}
\caption{Parameter tuning for vehicle tracking example. Left: state estimate
RMSE vs. number of iterations $\tilde k$. A grid search was performed to choose
$\lambda_x$ and $\lambda_y$ for each value of $\tilde k$. Right: contour plot of
state estimate RMSE \eqref{e-state-rmse} for different values of $\lambda_x$ and
$\lambda_y$.}
\label{f-vehicle-tuning}
\end{figure}

\PAR{Effect of parameters}

Figure \ref{f-vehicle-tuning} illustrates the effect of the parameters $\tilde
k$, $\lambda_x$, and $\lambda_y$ on the performance of the ISKF on the test
data. The left plot shows the best-achieved state estimate RMSE (by grid search)
for each value of $\tilde k$. The ISKF achieves a $30\%$ improvement over the KF
after two iterations, and does not improve after three iterations. The contour
plot on the right shows the state estimate RMSE for different values of
$\lambda_x$ and $\lambda_y$.

\subsection{Cascaded continuously-stirred tank reactor}\label{s-cstr-example}

\PAR{CSTR model}

The adiabetic continuous stirred-tank reactor (CSTR) is a commonly-appearing
system in the chemical process industry \cite{Bequette1998,SeborgEMD2016}. We
consider an ideal model of a single, first-order exothermic and irreversible
reaction taking place in a reactor tank, which is assumed to be perfectly-mixed.
The reagent enters the tank through the inlet at a constant rate, and the output
product leaves the reactor at the same constant rate. The state consists of the
concentration of the reagent $c$ and the temperature $\tau$ in the reactor, and
the dynamics are controlled by the inlet concentration $c^{\textrm{in}}$ and the
tank's jacket coolant temperature $\tau^{\textrm{c}}$.

The continuous-time dynamics are linearized around an operating point $(c_0,
\tau_0, c^{\textrm{in}}_0, \tau^{\textrm{c}}_0)$. In the linearized model, the
state is $\xi = (c - c_0, \tau - \tau_0)$, and the process is driven by $u =
(c^{\textrm{in}} - c^{\textrm{in}}_0, \tau^{\textrm{c}} - \tau^{\textrm{c}}_0)$.
We observe measurements of the reactor's temperature, but not of the reagent
concentration. The discrete-time dynamics with step size of $h$ are
\[
\xi_{t+1} = \tilde A \xi_t + \tilde B u_t, \quad y_t = \tilde C \xi_t + \nu_t,
\]
where $\nu_t\in\reals$ is the measurement noise and the matrices are
\cite{MathWorksCSTRModel2025}
\BEAS \tilde A &=& \BBM 1 - 5h + 4.33h^2 & -0.34h
+ 0.38h^2 \\ 
47.68h - 52.81h^2 & 1 + 2.79h - 4.29h^2
\EBM, \\
\tilde B &=& \BBM 
h -2.5h^2 & -0.05h^2 \\ 
23.84h^2 & 0.3h + 0.42h^2
\EBM, \\
\tilde C &=& \BBM 0 & 1 \EBM.
\EEAS
Note that here, $y_t$ represents a measurement of the temperature offset from
the operating point $\tau_0$, rather than the temperature itself.

\PAR{System model}

In this example, we consider a cascade of three such reactors, with the state of
each reactor being the input to the next. Let $c_i$ and $\tau_i$ be the reagent
concentration and temperature of the $i$-th reactor, respectively. Then, the
cascaded system has state $x\in\reals^6$ given by $x=(\xi_1, \xi_2, \xi_3)$,
where $\xi_i = (c_i - c_0, \tau_i - \tau_0)$ is the state of the $i$-th reactor.
The cascaded system has discrete-time model
\[
x_{t+1} = A x_t + w_t, \quad y_t = C x_t + v_t,
\]
where
\[
A = 
\BBM \tilde A & 0 & 0 \\
\tilde B & \tilde A & 0 \\
0 & \tilde B & \tilde A 
\EBM, \quad
C = \BBM \tilde C & 0 & 0 \\
0 & \tilde C & 0 \\
0 & 0 & \tilde C
\EBM,
\]
and $w\in\reals^6$ and $v\in\reals^3$ are the process and measurement noises,
with $F$ and $G$ matrices given by
\[
F = \frac{1}{\sqrt{10}}\BBM \tilde B & 0 & 0 \\
0 & \tilde B & 0 \\
0 & 0 & \tilde B
\EBM, \quad
G = \BBM 1 & 0 & 0 \\
0 & 1 & 0 \\
0 & 0 & 1
\EBM.
\]

The process noise $w_t$ and measurement noise $v_t$ are distributed according to
\[
w_t \sim \begin{cases}
\mathcal{N}(0, F F^T) & \text{with probability } 0.9, \\
\mathcal{N}(0, 100 F F^T) & \text{with probability } 0.1,
\end{cases}
\]
and
\[
v_t \sim \begin{cases}
\mathcal{N}(0, I) & \text{with probability } 0.9, \\
\mathcal{N}(0, 100 I) & \text{with probability } 0.1,
\end{cases}.        
\]
We discretized the continuous-time dynamics with a step size of $h=50
\textrm{ms}$. The operating point is $c_0 = 2\textrm{kmol/m}^3$, $\tau_0 =
373\textrm{K}$, $c^{\textrm{in}} = 10\textrm{kmol/m}^3$, and $\tau^{\textrm{c}}
= 299\textrm{K}$.

\begin{figure}
\centering
\includegraphics[width=0.8\textwidth]{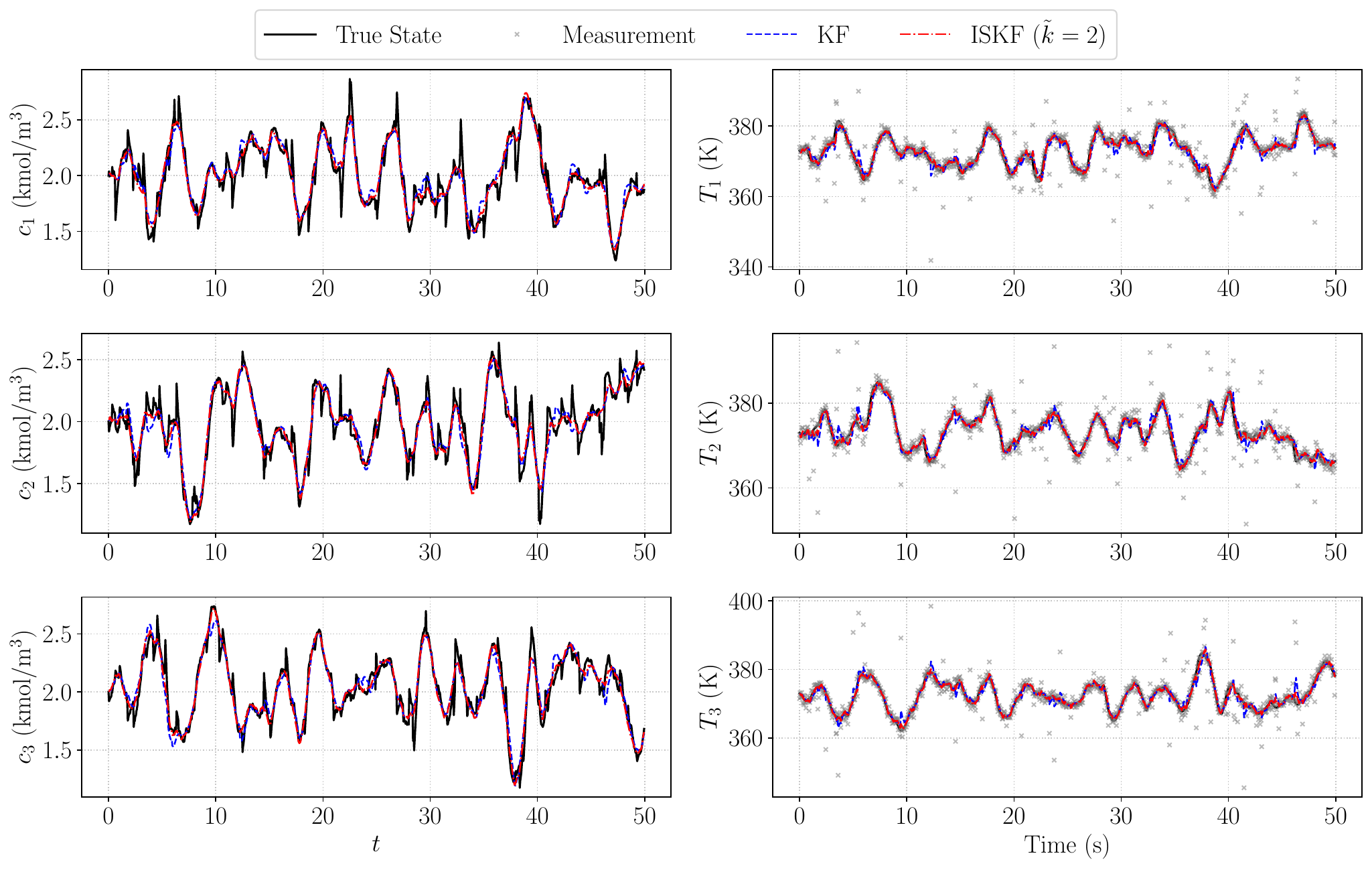}
\caption{True reagent concentration and temperature values over time, along with measurements and state estimates produced by the (steady-state) ISKF and KF.}
\label{f-cstr-trajectory}
\end{figure}

\begin{figure}
\centering
\includegraphics[width=0.95\textwidth]{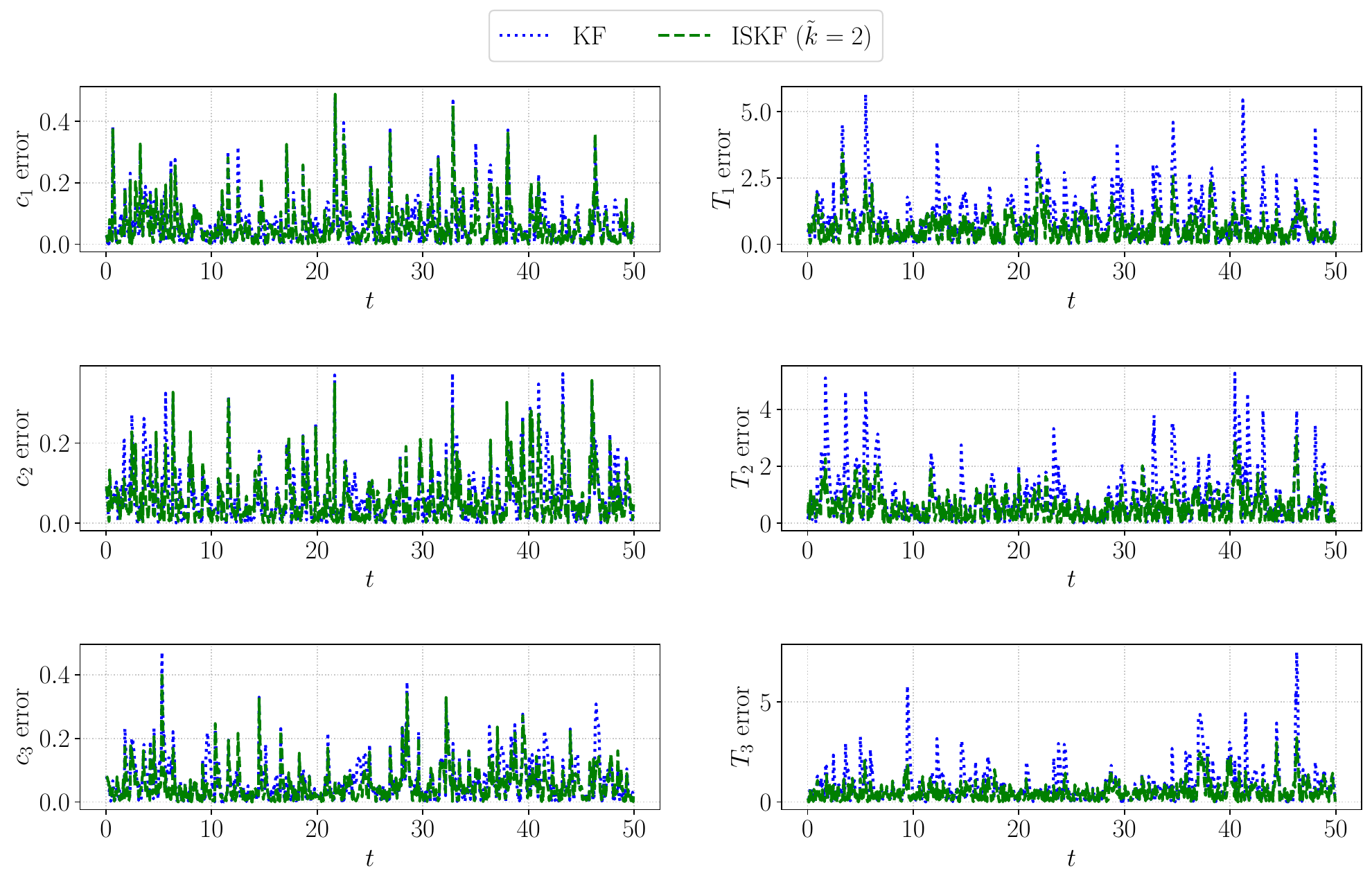}
\caption{State estimate errors (absolute values) for the KF and the two-iteration ISKF.}
\label{f-cstr-errors}
\end{figure}
    
\PAR{Performance comparison}

With the number of iterations fixed at $\tilde k=2$, we selected the parameters
to be
\[
\lambda_x = 0.10, \quad \lambda_y = 3.3
\]
using the grid search procedure described in \S\ref{s-parameter-selection}. Like
in the vehicle tracking example, the grid search was carried out over the
predicted measurement RMSE \eqref{e-meas-rmse} on a separate simulated
trajectory of measurements. Figure \ref{f-cstr-trajectory} shows the true
reagent concentration and temperature values over time, along with measurements
and estimates produced by the (steady-state) ISKF and KF. Figure
\ref{f-cstr-errors} shows the state estimate errors (absolute values) for the KF
and the two-iteration ISKF.

Table \ref{t-cstr-performance} shows the state estimate RMSE
\eqref{e-state-rmse} evaluated for several filters on the same test trajectory.
The ISKF with $\tilde k=1$ achieves comparable performance to WoLF, as both are
only designed to reject measurement outliers.

\begin{figure}
\caption{Parameter tuning for CSTR example. Left: state estimate
RMSE vs. number of iterations $\tilde k$. A grid search was performed to choose
$\lambda_x$ and $\lambda_y$ for each value of $\tilde k$. Right: contour plot of
state estimate RMSE \eqref{e-state-rmse} for different values of $\lambda_x$ and
$\lambda_y$.}
\label{f-cstr-tuning}
\centering
\begin{subfigure}{0.45\textwidth}
\centering
\includegraphics[width=\textwidth]{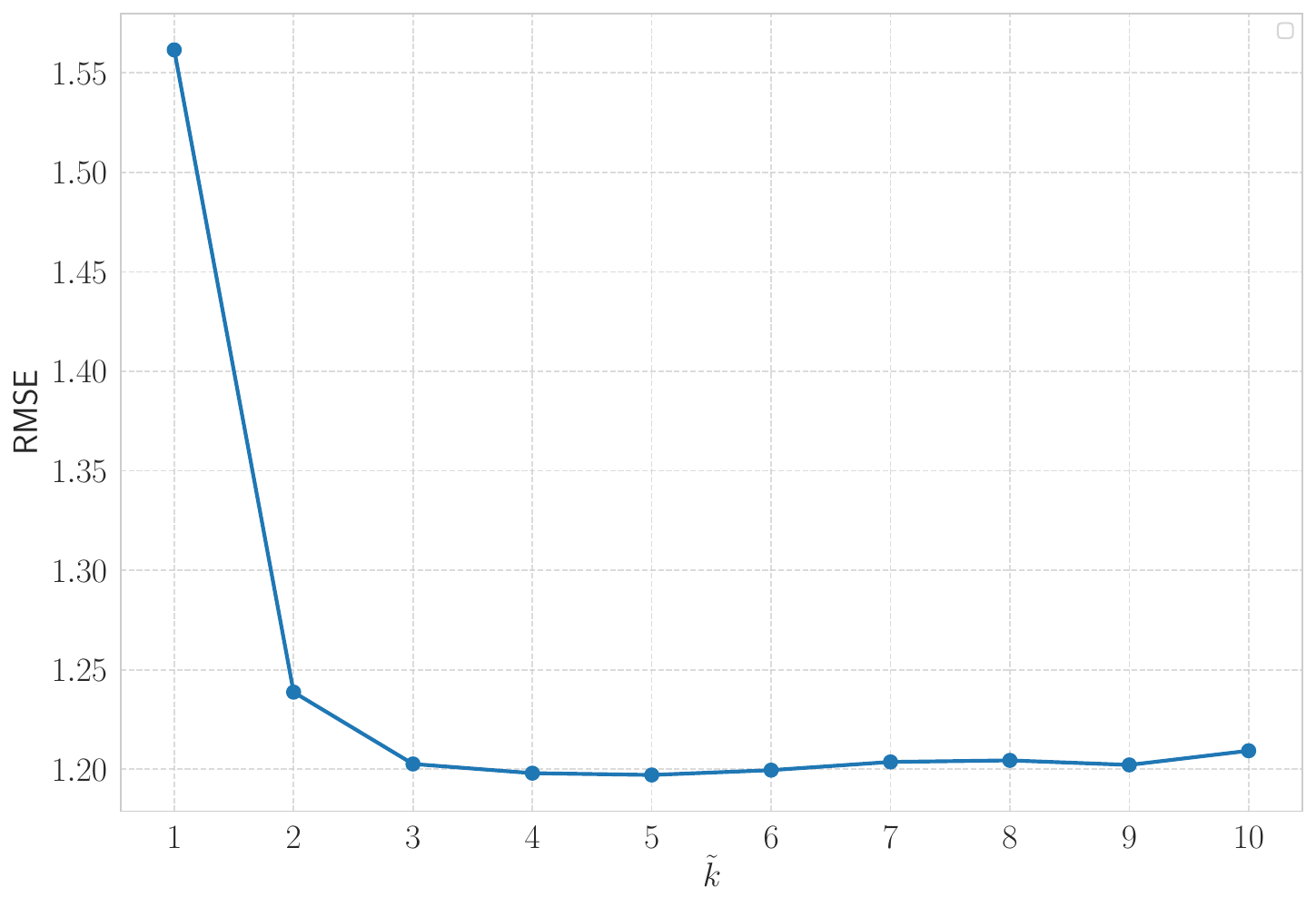}
\caption{\label{f-cstr-iter-sweep}}
\end{subfigure}
\begin{subfigure}{0.45\textwidth}
\centering
\includegraphics[width=\textwidth]{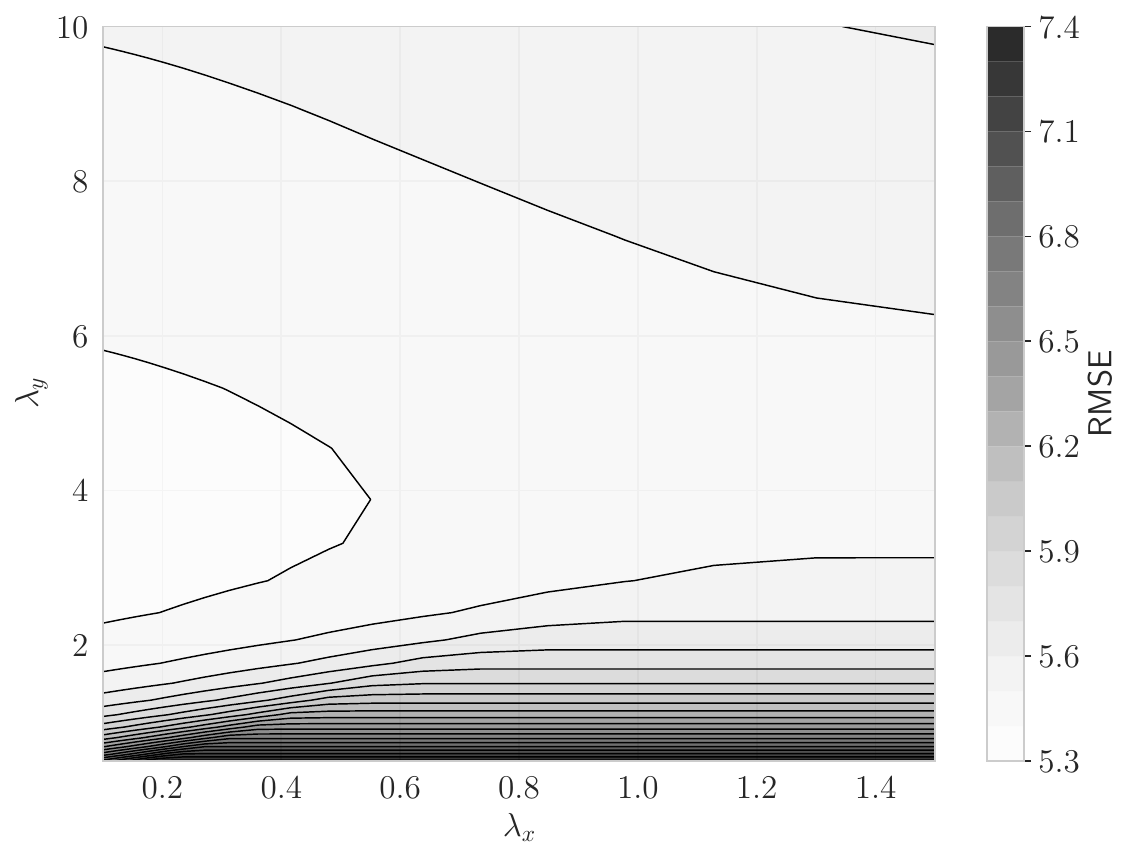}
\caption{\label{f-cstr-contour}}
\end{subfigure}
\end{figure}

\begin{table}[b]
\caption{State estimate RMSE comparison of several filtering methods for the
cascaded CSTR example. All filters were tuned using the same data, and
evaluated on the same test data.}
\label{t-cstr-performance}
\centering
\begin{tabular}{|c||c|c|}
\hline
Method & RMSE & Improvement over KF \\
\hline
\hline
KF & 2.46 & - \\
ISKF ($\tilde k = 1$) & 1.57 & 36\% \\
ISKF ($\tilde k = 2$) & 1.25 & 49\% \\
ISKF ($\tilde k = 3$) & 1.21 & 51\% \\
WoLF & 1.79 & 27\% \\
Huber KF & 1.56 & 37\% \\
\hline
\end{tabular}
\end{table}

\begin{table}[b!]
\centering
\caption{Results of tuning the step size $\eta$ jointly with $\lambda_x$ and
$\lambda_y$. RMSE is evaluated on the same test data as in
\S\ref{s-vehicle-example} and \S\ref{s-cstr-example}. The RMSE with no outliers is
computed by removing the outliers from the test data, so that the process and
measurement noises are Gaussian with fixed covariance.}
\label{t-tuning-step-size}
\begin{tabular}{|c|c|c|c||c|c|}
\hline
Method & $\eta$ & $\lambda_x$ & $\lambda_y$ & RMSE & RMSE (no outliers)\\
\hline
\hline
KF & - & - & - & 4.55 & 1.40 \\
\hline
ISKF ($\tilde k=2$) & 1 & 0.10 & 1.83 & 3.19 & 1.50 \\
ISKF ($\tilde k=2$) & 2.64 & 0.10 & 0.89 & 3.15 & 1.67 \\
\hline
\end{tabular}
\vspace{0.5em}
\subcaption{Vehicle tracking example}
\vspace{0.5em}
\begin{tabular}{|c|c|c|c||c|c|}
\hline
Method & $\eta$ & $\lambda_x$ & $\lambda_y$ & RMSE & RMSE (no outliers)\\
\hline
\hline
KF & - & - & - & 2.46 & 0.68 \\
\hline
ISKF ($\tilde k=2$) & 1 & 0.10 & 3.80 & 1.27 & 0.78 \\
ISKF ($\tilde k=2$) & 1.83 & 0.26 & 2.34 & 1.23 & 0.92 \\
\hline
\end{tabular}
\vspace{0.5em}
\subcaption{CSTR example}
\end{table}

\PAR{Effect of parameters}

Figure \ref{f-cstr-tuning} illustrates the effect of the parameters $\tilde k$,
$\lambda_x$, and $\lambda_y$ on the performance of the ISKF on the test data.
The left plot shows the best-achieved state estimate RMSE (by grid search) for
each value of $\tilde k$. The ISKF achieves a $49\%$ improvement over the KF
after two iterations, and, like in the vehicle tracking example, does not
improve after three iterations. The contour plot on the right shows the state
estimate RMSE for different values of $\lambda_x$ and $\lambda_y$.

\subsection{Tuning step size}

As discussed in \S\ref{s-parameter-selection}, the step size $\eta$ may be tuned
jointly with $\lambda_x$ and $\lambda_y$ as part of the same grid search
procedure. For both the vehicle tracking and CSTR examples, we found that while
increasing $\eta$ can provide marginal improvements in performance, the
resulting filter underperforms when there are no outliers in the simulation,
\ie, the process noise and measurement noises are Gaussian with fixed
covariance. In the following, we consider a grid search over $20$ values of
$\eta$ between 0.1 and 100, logarithmically-spaced. The results are shown in
Table \ref{t-tuning-step-size}. In the vehicle tracking example, the (two-step)
ISKF with a tuned value of $\eta=2.64$ achieves a $1\%$ improvement over the
ISKF with $\eta=1$ on the test data, but underperforms by $10\%$ when the
outliers are removed from the test data. In the CSTR example, the ISKF with
tuned value $\eta=1.83$ achieves a $3\%$ improvement over the ISKF with $\eta=1$
on the test data, but underperforms by $15\%$ when the outliers are removed from
the test data.

\section{Conclusion}\label{s-conclusion}

We have introduced the iteratively saturated Kalman filter, a modification of
the standard KF's update step that makes it robust to outliers. The method is
derived as a scaled gradient method for solving a particular convex maximum a
posteriori estimation problem.  The steady-state variant of the ISKF matches the
computational efficiency of the steady-state KF, and is well-suited for
real-time applications.

\backmatter

\printbibliography
\printindex

\chapter{About the Authors}

\noindent\begin{wrapfigure}[8]{l}{0.4\textwidth}
\vspace{-\baselineskip}
\centering
\includegraphics[width=\linewidth,keepaspectratio]{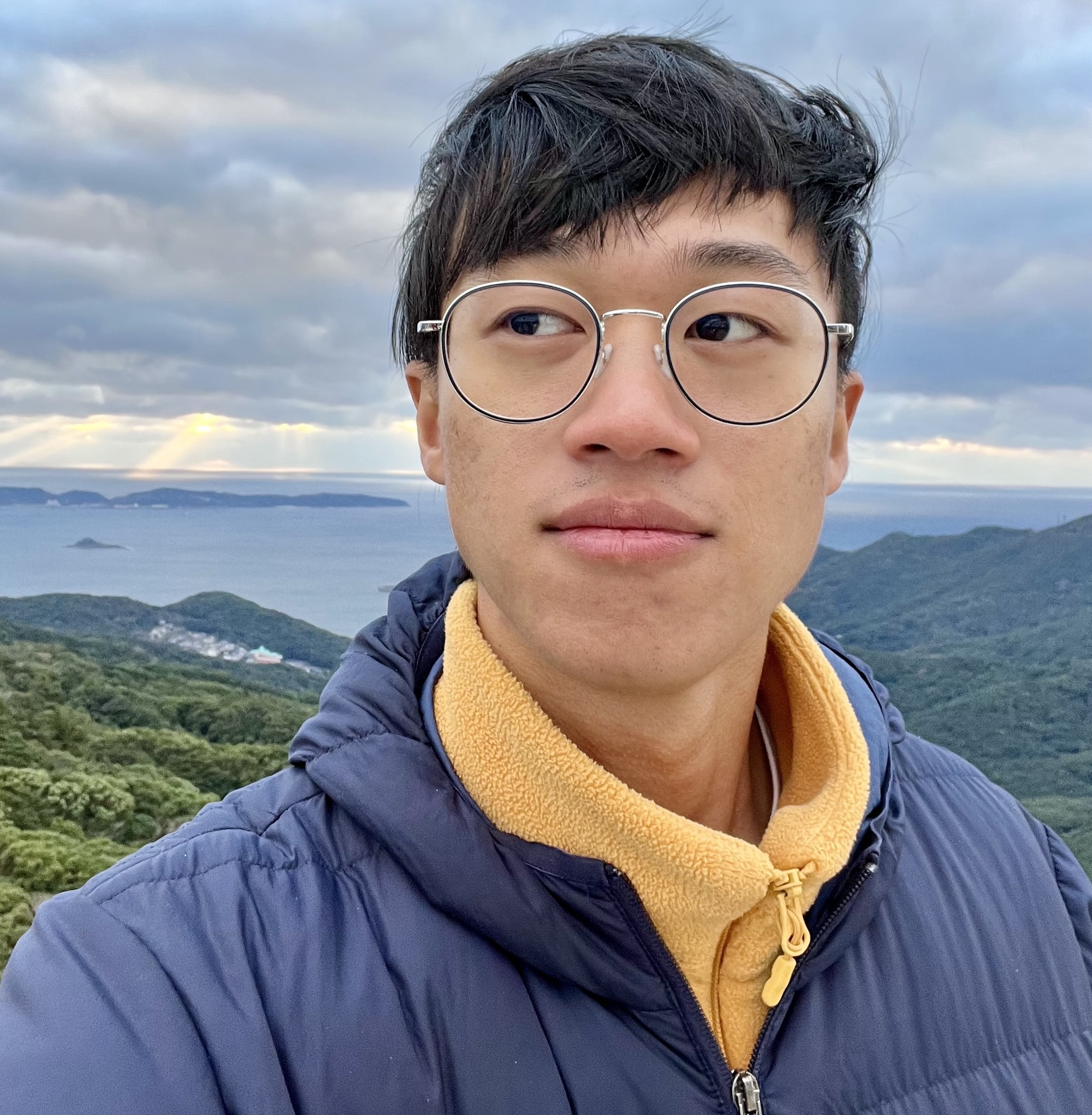}
\end{wrapfigure}
Alan Yang is a Ph.D candidate in Electrical Engineering at Stanford University.
He received the B.S. degree in Electrical Engineering from the University of
Illinois at Urbana-Champaign in 2018. His research interests include convex
optimization and machine learning, control systems, and signal processing.

\par
\mbox{}
\par
\vspace{5\baselineskip}

\noindent\begin{wrapfigure}[7]{l}{0.4\textwidth}
\vspace{-\baselineskip}
\centering
\includegraphics[width=\linewidth,keepaspectratio]{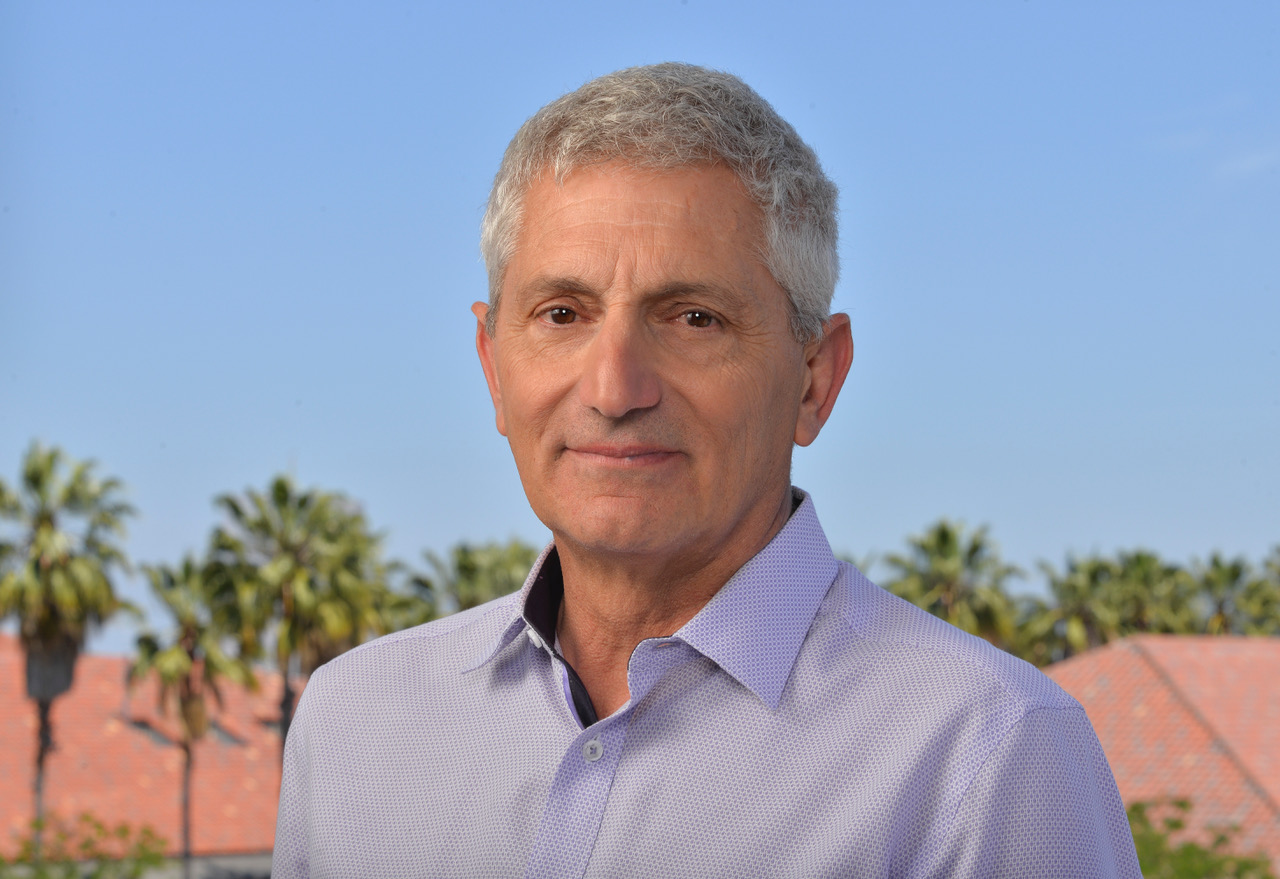}
\end{wrapfigure}
Stephen Boyd is the Samsung Professor of Engineering, and Professor of
Electrical Engineering at Stanford University. He received the A.B. degree in
Mathematics from Harvard University in 1980, and the Ph.D. in Electrical
Engineering and Computer Science from the University of California, Berkeley, in
1985, before joining the faculty at Stanford. His current research focus is on
convex optimization applications in control, signal processing, machine
learning, and finance. He is a member of the US National Academy of Engineering,
a foreign member of the Chinese Academy of Engineering, and a foreign member of
the National Academy of Korea.

\end{document}